\documentclass[aip,twocolumn,showpacs,superscriptaddress,reprint,amsmath,amssymb,floatfix]{revtex4-1}
\pdfoutput=1
\usepackage{graphicx}
\usepackage{dcolumn}
\usepackage{bm}
\usepackage{amssymb}
\usepackage{epsfig}
\usepackage{pdfsync}
\usepackage[colorlinks,linkcolor=blue,citecolor=blue,urlcolor=blue]{hyperref}

\begin{document}
\title{Mesoscopic Josephson junctions with switchable current-phase relation}
\author{E. Strambini}
\email{e.strambini@sns.it}
\affiliation{NEST Istituto Nanoscienze-CNR  and Scuola Normale Superiore, I-56127 Pisa, Italy}
\author{F. S. Bergeret}
\email{sebastian\_bergeret@ehu.es}
\affiliation{Centro de F\'{i}sica de Materiales (CFM-MPC), Centro Mixto CSIC-UPV/EHU, Manuel de Lardizabal 4, E-20018 San Sebasti\'{a}n, Spain}
\affiliation{Donostia International Physics Center (DIPC), Manuel de Lardizabal 5, E-20018 San Sebasti\'{a}n, Spain}
\affiliation{Institut f\"ur Physik, Carl von Ossietzky Universit\"at, D-26111 Oldenburg, Germany}
\author{F. Giazotto}
\email{f.giazotto@sns.it}
\affiliation{NEST Istituto Nanoscienze-CNR  and Scuola Normale Superiore, I-56127 Pisa, Italy}
\begin{abstract}
We propose and analyze a mesoscopic Josephson junction consisting of two ferromagnetic insulator-superconductors (FI-Ss) coupled through  a normal metal (N) layer. 
The Josephson current of the junction is non-trivially affected by the spin-splitting field induced by the FIs in the two superconductors. In particular,  it shows sizeable enhancement by increasing the amplitude of the exchange field ($h_{ex}$) and displays a switchable current-phase relation which depends on the relative orientation of $h_{ex}$ in the FIs. In a realistic EuS/Al-based setup this junction can be exploited as a high-resolution threshold sensor for the  magnetic field as well as an on-demand tunable kinetic inductor. 
\end{abstract}
\maketitle
\begin{figure}[b!]
\includegraphics[width=\columnwidth]{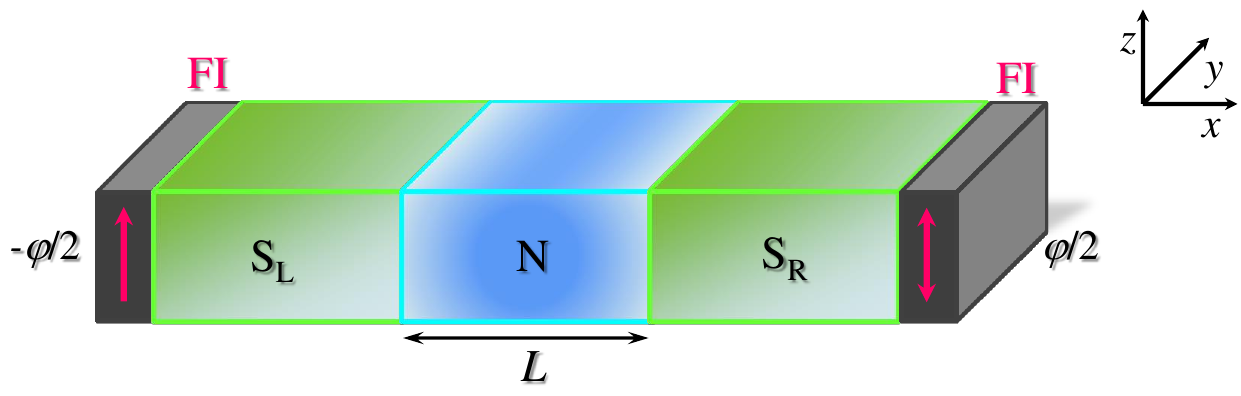}
\caption{\label{fig1} 
Scheme of the FIS-N-FIS Josephson junction. The red arrows indicate the direction of the magnetization in the ferromagnetic insulators (FIs).
N stands for a conventional normal metal whereas $\varphi$ is the  quantum phase difference over the junction. $L$ denotes the length of the weak link.
}
\end{figure}
The interplay between superconductivity and ferromagnetism in superconductor-ferromagnet (SF) hybrids exhibits a large variety of effects studied along the last years \cite{Bergeret_Odd_2005,Buzdin_Proximity_2005}.
Experimental research mainly focuses on the control of the 0-$\pi$ transition in the S-F-S junctions~\cite{Ryazanov_coupling_2001} and on the creation, detection, and manipulation of triplet correlations in SF hybrids~\cite{Bergeret_long-range_2001,Keizer_spin_2006,robinson_controlled_2010,Khaire_observation_2010,Klose_optimization_2012,Anwar_Long-range_2010}. 
From a fundamental point of view, the key phenomenon for the understanding of these effects is the \emph{proximity effect} in SF hybrids, and how the interplay between superconducting and magnetic correlations affects their thermodynamic and transport properties.
While most of theoretical and experimental investigations on SF structures deal mainly with the penetration of superconducting correlations into the F regions, it is also widely known that magnetic correlations can be induced in the superconductor via the inverse proximity effect~\cite{Tokuyasu_Proximity_1988,Bergeret_induced_2004,Bergeret_inverse_2005,Xia_Inverse_2009}. 
If the ferromagnet is an insulator (FI), on the one hand,  superconducting correlations are weakly suppressed at the FI-S interface and, on the other hand,  a finite exchange field ($h_{ex}$) is induced at the interface and penetrates into the S  region over distances of the order of the coherence length ~\cite{Tokuyasu_Proximity_1988}.
This results in a spin splitting of the density of states (DoS) of the superconductor, as observed in a number of experiments~\cite{hao_spin-filter_1990,santos_determining_2008,Catelani_Fermi-liquid_2008,Xiong_Spin-resolved_2011}. 
This  spin-splitting may lead to interesting phenomena such as the absolute spin valve effect~\cite{Meservey_Spin-polarized_1994,huertas-hernando_absolute_2002,Giazotto_superconductors_2008}, the magnetothermal Josephson valve~\cite{giazotto_phase-tunable_2013,bergeret_Phase-dependent_2013}, and the enhancement of the Josephson current in SF-I-SF junctions (I stands for a conventional insulator)~\cite{bergeret_enhancement_2001,krivoruchko_inversion_2001,chtchelkatchev_josephson_2002,zaitsev_peculiarities_2009,robinson_enhanced_2010}. 

In this Letter we investigate the Josephson current in a mesoscopic FIS-N-FIS junction.  As we show below, the presence of a normal metal 
instead of an insulator  provides access to a rich nontrivial phenomenology which stems from the interplay of  phase-tunable superconducting and magnetic
 correlations in the N region.  
The impact of magnetic correlations on the Josephson coupling is explored for two transversal magnetization directions of the FIS layers, i.e., parallel (P) and antiparallel (AP). This results in an enhancement of the critical current  and in peculiar current-phase relations (CPRs). 
We propose realistic hybrid setups to observe these anomalous effects, and discuss how to exploit them for  practical applications.

The Josephson current in the structure shown in Fig.~\ref{fig1} is calculated with the help of  the quasiclassical Green's functions formalism. 
The S leads, being in contact with a FI, show a spin-splitting in the DoS. 
By assuming that both electrodes are identical,  the GFs describing the left (right) L(R)  electrodes are given by:
$$
\check G^{R(L)}=\hat G^{R(L)}\tau_3+\hat F^{R(L)}\left[\cos(\varphi/2)i\tau_2\pm\sin(\varphi/2)i\tau_1\right],
$$
where $\tau_{1,2,3}$ are the Pauli matrices, $\varphi$ is the phase difference between the electrodes, and 
$$
\hat G^{R(L)}=\left( 
\begin{array}{cc}
 G^{R(L)}_+ &  0 \\
 0 &  G^{R(L)}_-
 \end{array}
 \right), \;
\hat F^{R(L)}=\left( 
\begin{array}{cc}
 F^{R(L)}_+ &  0 \\
 0 &  F^{R(L)}_-
 \end{array}
 \right).
$$
Here,  the subindex $\pm$ denotes the spin index with respect to the local exchange field $h_{ex}$, i.e., $G^{R(L)}_\pm=(\omega_n\pm ih_{ex}^{R(L)})/\sqrt{(\omega_n\pm ih_{ex}^{R(L)})^2+\Delta^2}$ and $F^{R(L)}_\pm=\Delta/\sqrt{(\omega_n\pm ih_{ex}^{R(L)})^2+\Delta^2}$, where $\omega_n=\pi k_B T(2n+1)$ is the Matsubara frequency, $T$ is the temperature, and $k_B$ is the Boltzmann constant. $\Delta(h_{ex},T)$ is the effective superconducting order parameter calculated self-consistently from the BCS gap equation~\cite{giazotto_quantum_2013}. In the N region the GFs have to be determined by solving the Usadel equation, 
$D\partial_x(\check g\partial_x \check g)+\omega_n[\tau_3,\check g]=0$, 
where $D$ is the diffusion coefficient.  At the boundaries with the FIS electrodes ($x=0,L$) the function $\check g$ obeys the Kupriyanov-Lukichev  condition
$\check g\partial_x\check g|_{x=0,L}=\pm \gamma[\check g, \check G^{R(L)}]$.
Here $\gamma= 1/(2R_b\sigma)$, where $R_b$ is the contact resistance per unit area of the FIS-N interface, and $\sigma$ is the conductivity of the N region. 
 The Josephson current through the junction is then obtained from the expression
\begin{equation}
\label{current}
J=\frac{i\pi\sigma}{2e} T\sum_{\omega_n}{\rm Tr}\tau_3\check g\partial_x\check g.
\end{equation}
In order to provide useful compact analytical expressions for the Josephson current,  we mainly focus our analysis  on  two limiting cases: First, we consider a N region of  arbitrary length $L$ which is weakly coupled to the FI-S electrodes; Second,  we focus on arbitrary interface resistance but a short N region. 
 
{\it Weak-coupling limit.}
If the interface resistance $R_b$ is large enough, the proximity effect in N is weak and $\check g$ can be approximated by $\check g\approx {\rm sgn}\omega \tau_3+\check f$, where $\check f$ is the anomalous GF induced in the N region.   
The linearized Usadel equation can be solved easily and  from this solution one obtains the expression for the Josephson current. 
We distinguish two magnetic configurations of the FIS electrodes. In the parallel (P) case we get
\begin{equation}
\label{JPlin}
J_P=\frac{\pi \gamma}{eR_b} T\sum_{\omega_n}\frac{1}{\kappa_\omega\sinh(L\kappa_\omega)}\left(F_+^2+F_-^2\right)\sin\varphi ,
\end{equation}
where $\kappa_\omega=\sqrt{2|\omega|/\hbar D}$.
In the case that the magnetizations of the FIS electrodes are arranged in the antiparallel  (AP) configuration we obtain
\begin{equation}
\label{JAPlin}
J_{AP}=\frac{2\pi \gamma}{eR_b} T\sum_{\omega_n}\frac{1}{\kappa_\omega\sinh(L\kappa_\omega)}F_+F_-\sin\varphi . 
\end{equation}

\begin{figure}[th!]
\includegraphics[width=\columnwidth]{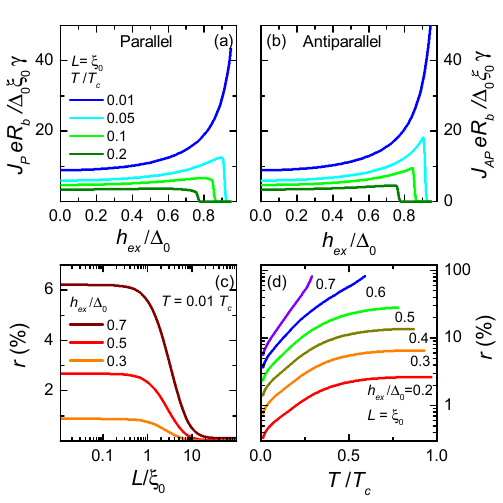}
\caption{
\label{fig2} 
Dependence of the critical current on the exchange field ($h_{ex}$) for the P (a) and AP (b) configurations at a few temperatures. 
Length and temperature dependence of $r$, (c) and (d), respectively, for different values of $h_{ex}$. All the curves are calculated at $\varphi = \pi /2$. $T_c$ denotes the zero-exchange field superconducting critical temperature, $\xi_0=\sqrt{\hbar D/\Delta_0}$ is the superconducting coherence length, and $\Delta_0$ is the zero-temperature, zero-exchange field energy gap.
}
\end{figure}

From Eqs. (\ref{JPlin}) and (\ref{JAPlin}) it follows that the Josephson CPR is \emph{sinusoidal}, as  expected in the weak-coupling limit, while the amplitude of the current shows nontrivial features in the presence of $h_{ex}$, as displayed in Fig.~\ref{fig2}. 
At low temperature $J$ shows a sizeable enhancement in $h_{ex}$ both in the P and AP configuration [see Fig.~\ref{fig2}(a) and (b), respectively~\cite{bergeret_footntefisnfis_2014}]. 
While in the AP configuration this enhancement is in agreement with the one expected for SF-I-SF junctions~\cite{bergeret_enhancement_2001,chtchelkatchev_josephson_2002,golubov_current-phase_2004}, 
the enhancement of $J_P$ is at a first glance unexpected, since usually this configuration is characterized by a suppression of the Josephson current by increasing $h_{ex}$.
We attribute the origin of this enhancement to a strong Josephson coupling originating at the FIS-N interfaces, 
similarly to what has been reported for  FIS-I-S structures  ~\cite{Bergeret_Manifestation_2014} and for Josephson junctions between superconductors with intrinsic exchange fields\cite{chtchelkatchev_josephson_2002}.  
The enhancement is more pronounced  at low $T$ when the order parameter of the FIS bilayer is only weakly affected by $h_{ex}$. 
By increasing the temperature  ($T \gtrsim 0.1 T_c$, where $T_c$ is the superconducting critical temperature at $h_{ex}=0$) the order parameter is strongly suppressed, and the enhancement reduced.

This picture is confirmed by the different behavior of the two Josephson currents in $L$.
In Fig.~\ref{fig2}(c) we plot the relative difference between $J_P$ and $J_{AP}$, quantified by the coefficient $r = (J_{AP} - J_{P})/J_P$, as a function of $L$. 
For  $L\gg \xi_0=\sqrt{\hbar D/\Delta_0}$ the two condensates in the FISs are fully decoupled, and the local FIS-N coupling is mainly affecting $J$ which is thus independent of the magnetic configuration (i.e., $r \simeq 0$). Only for $L \lesssim  \xi_0$ the AP configuration shows a stronger enhancement of the supercurrent due to the additional coupling between the two FISs.  
Figure ~\ref{fig2}(d) shows several curves characterizing the $T$ dependence of $r$ at different $h_{ex}$ and for $L=\xi_0$. 
As it can be deduced from Eqs.~(\ref{JPlin})and~(\ref{JAPlin}) at low $T$ or low $h_{ex}$ the two Josephson currents are similar, than $r\simeq 0$ and the orientation-independent coupling described above is mainly driving the behavior of the Josephson junction. 
Only for $h_{ex}\gtrsim0.5 \Delta_0$ and $T\gtrsim 0.1 T_c$ the difference between $J_P$ and $J_{AP}$ becomes relevant, and the magnetic configuration start playing an important role. 

All the above results are valid when the transparency of the FIS-N interface is low. As a second limiting case  we 
investigate the influence of the transparency on the critical current in the short junction limit, i. e., when the length  of the  junction is smaller than the characteristic penetration length of the superconducting correlations.

{\it Short junction limit.}
If we now assume  $L\ll \sqrt{\hbar D / k_B T_c}$ 
 we can integrate the Usadel equation over the thickness by using the Kupriyanov-Lukichev condition (see, for instance, Ref.~\cite{strambini_proximity_2014}). 
 In this case we obtain for the Josephson current in the P configuration
\begin{equation}
\label{JPshort}
J^{P}=\frac{\pi T }{eR_b}
\sum_{\omega_n}{\rm Re}\left[\frac{ F^2_+ \sin(\varphi)}{\sqrt{(2G_++\omega_n /\epsilon_b )^2+4F_+^2\cos^2(\varphi/2)}}\right],
\end{equation}
where $\epsilon_b=\hbar D\gamma/L $.
In the AP case we obtain 
\begin{equation}
\label{JAPshort}
J^{AP}=\frac{\pi T}{eR_b}\sum_{\omega_n}\frac{F_+F_- \sin(\varphi)}{\sqrt{\left[\mathcal{G}+\omega_n/ \epsilon_b\right]^2+\left[\mathcal{F}\cos\varphi+2F_+F_-\right]}},
\end{equation}
where $\mathcal{G}= G_++G_-$ and $\mathcal{F}= F_+^2+F_-^2$.
\begin{figure}[th!]
\includegraphics[width=\columnwidth]{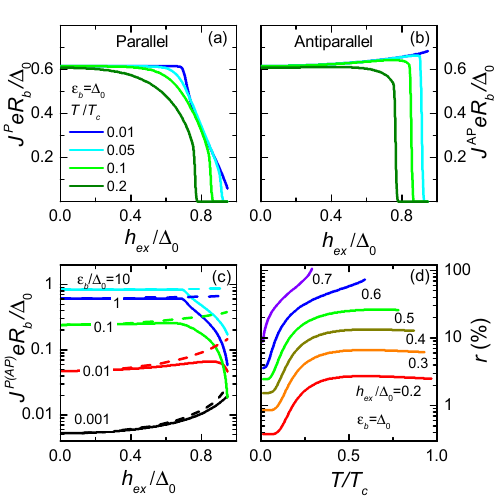}
\caption{
\label{fig3} 
Dependence of the critical current on $h_{ex}$ for the P (a) and AP (b) configuration calculated for few temperatures and $\varepsilon_b=\Delta_0$. (c) Comparison between $J^P$ (solid line) and $J^{AP}$ (dashed line) vs $h_{ex}$ for different FIS-N coupling $\epsilon_b$ at $T = 0.01 T_c$.
(d) Temperature dependence of $r$ for different values of $h_{ex}$. All the curves are calculated at $\varphi = \pi /2$.
}
\end{figure}
\begin{figure}[th!]
\includegraphics[width=\columnwidth]{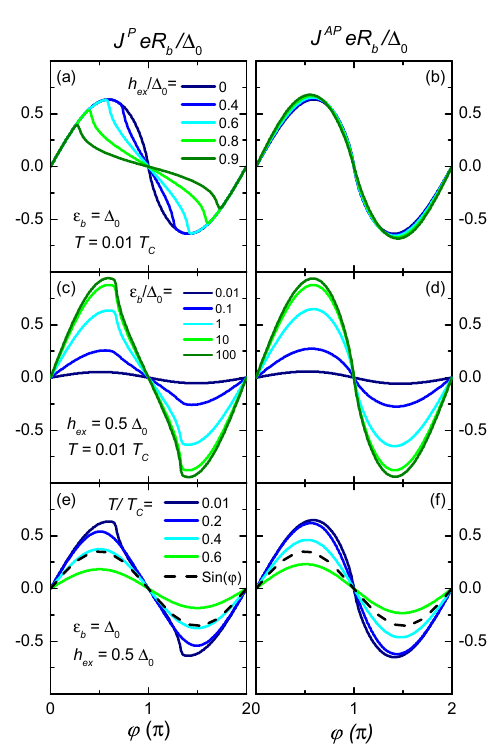}
\caption{
\label{fig4} 
Comparison between the CPRs of the P and AP configurations for different $h_{ex}$, (a) and (b), $\epsilon_b$, (c) and (d), and temperatures, (e) and (f).
}
\end{figure}

In Fig.~\ref{fig3}(a) and (b) we show the dependence of $J^P$ and $J^{AP}$, respectively, on $h_{ex}$. 
Unlike the weak-coupling limit (see Fig.~\ref{fig2}), for $\epsilon_b \gtrsim \Delta_0$ the enhancement of $J$ in $h_{ex}$ is strongly reduced in the AP configuration, and almost negligible in the P one. 
In agreement with the results obtained in the weak-coupling regime, the Josephson current enhancement is fully recovered at very low transparency, i.e., for $\epsilon_b \lesssim 0.001 \Delta_0$  [see Fig.~\ref{fig3}(c)].
This trend confirms the crucial role played by the discontinuity of the condensate at the FIS-N interface in determining the properties of  $J$: the discontinuity is indeed smoothed at high interface transparency when the coupling between the two FISs becomes relevant.

Analogously to the  weak-coupling limit, the difference between $J^{P}$ and $J^{AP}$ is quantified by  the coefficient $r$ plotted in Fig.~\ref{fig3}(d) vs $T$ for a few $h_{ex}$ values. 
In this approximation $r$ is zero only for the trivial condition $h_{ex}=0$, while at low $T$ it saturates demonstrating that for finite FIS-N transparency 
the correlations between the two FISs are small ($r < 10 \%$) but still present even at very low temperature ($T \ll T_c$).
These correlations become more relevant ($r \sim 100 \%$) at higher $T$ ($\gtrsim 0.1 T_c$) and $h_{ex}$ ($\gtrsim0.5\Delta_0$), similarly to the weak-coupling limit [see Fig.~\ref{fig2}(d)].

The difference between $J^{P}$ and $J^{AP}$ is even more apparent in the Josephson CPR of the two magnetic configurations. 
As it can be noted from Eqs.~(\ref{JPshort}) and~(\ref{JAPshort}) the CPRs are strongly deviating from the usual sinusoidal behavior  typical of the weak-coupling limit. 
Such a  deviation, which stems from the enhanced coupling between the condensate and N, is much more pronounced in the P configuration [see Fig.~\ref{fig4}(a)] whereas in the AP one [see Fig.~\ref{fig4}(b)] $h_{ex}$ only weakly affects the CPR.
We emphasize that the behavior of $J^{P}(\varphi)$ in the presence of a finite $h_{ex}$ is similar to the CPR  of SFS junctions~\cite{golubov_nonsinusoidal_2002,Golubov_current_2005},  for which a $\pi $ shift is obtained  if  $h_{ex}$ is sufficiently large.
Theoretically, a similar  $\pi$ shift is present in our junctions as well, but only for  $h_{ex} \gtrsim 0.95 \Delta_0$.

The non-sinusoidal critical current  behavior is furthermore enhanced by increasing the transparency of the FIS-N interfaces (i.e., for large $\epsilon_b$ values) and by lowering the temperature, as displayed in Fig.~\ref{fig4}. 
The anomalous shape of the CPR in the P configuration is  of particular relevance since the damping of the slope around $\pi$  results in a strong increase of the Josephson junction kinetic inductance [$\propto (\partial J / \partial \varphi)^{-1}$] in a region of phases ($\varphi \sim \pi$) very important for several phase-controlled devices~\cite{giazotto_superconducting_2010,meschke_tunnel_2011,valentini_andreev_2014,strambini_proximity_2014,ronzani_highly_2014,jabdaraghi_non-hysteretic_2014}.

\begin{figure}[t!]
\includegraphics[width=\columnwidth]{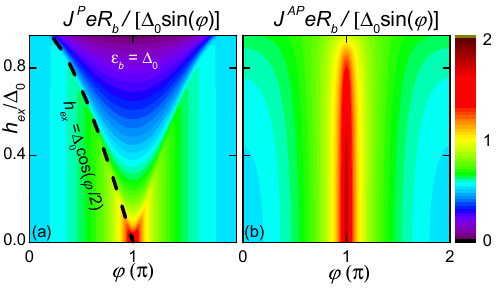}
\caption{
\label{fig5} 
Color plots of the deviation of the Josephson current from the sinusoidal CPR calculated in the P (a) and AP (b) configuration vs $\varphi$ and $h_{ex}$. In panel (a) the dashed line corresponds to the curve $h_{ex}= |\Delta_0 \cos(\varphi/2)|$. All the calculations were performed at $T = 0.01 T_c$ and for $\varepsilon_b=\Delta_0$.
}
\end{figure}

In order to quantify  the relative deviation of the CPR from the sinusoidal behavior we plot in Fig.~\ref{fig5} the ratio between the Josephson currents and the $\sin(\varphi)$ vs $h_{ex}$. 
For $h_{ex}=0$ this function has a maximum at $\varphi = \pi$ corresponding to the skewing observed  in transparent short SNS junctions~\cite{golubov_current-phase_2004}. 
In the P configuration, this maximum splits in $h_{ex}$ according to the relation $h_{ex}= |\Delta_0 \cos(\varphi/2)|$ [see the dashed line in Fig.~\ref{fig5}(a)] which corresponds to the matching between the exchange energy and the minigap induced in the N region. By contrast, in the AP configuration the peak appears always at $\varphi=\pi$,  and is only weakly damped by increasing $h_{ex}$.

Due to the request of high $h_{ex}$ ($\lesssim \Delta_0$), suitable candidate materials for the implementation of the FIS-N-FIS junction are europium (Eu) chalcogenides (like EuO~\cite{santos_determining_2008}, EuS~\cite{hao_spin-filter_1990} or EuSe~\cite{moodera_variation_1993}) for the FI combined with superconducting aluminum (Al), and copper~\cite{giazotto_superconducting_2010,ronzani_highly_2014} or silver~\cite{le_Sueur_phase_2008} for the N region. 
The Eu chalcogenides can indeed induce very large values of $h_{ex}$ which, depending on  the quality of the FI-S interface, lead to a Zeeman splitting tunable from 0.1 meV up to few meV, energies comparable to the Al superconducting gap ($\sim 0.2$~meV). 
Moreover, in EuSe films $h_{ex}$  can be tuned  by applying an external magnetic field. This  makes EuSe  suitable for highly-sensitive devices.
In order to operate at higher temperatures ($> T_c$(Al)~$\sim 1 $~K), gadolinium nitride/niobium nitride (GdN/NbN) bilayers~\cite{senapati_spin-filter_2011,Pal_pure_2014} can be an alternative choice to the above proposed materials since they can induce a similar $h_{ex}$ but with a higher critical temperature ($T_c \simeq 15 $~K).
With these materials it would be possible to realize the unconventional FIS-N-FIS Josephson junctions described above aiming at the implementation of nanoscale circuitry with novel functionalities, including vectorial threshold sensors for the magnetic field as well as tunable kinetic inductors.
In this latter case, the anomalous CPRs shown in Fig.~\ref{fig4}(a) demonstrate that the kinetic inductance of the Josephson weak link can be tuned by $h_{ex}$, and maximized around $\sim\pi$. This is indeed essential to provide a robust and well-defined phase drop across the junction in a number of nanodevices~\cite{giazotto_superconducting_2010,meschke_tunnel_2011,valentini_andreev_2014,strambini_proximity_2014,ronzani_highly_2014,jabdaraghi_non-hysteretic_2014}.

The work of E.S. and F.G. was partially funded by the European Research Council under the European Union's Seventh Framework Programme (FP7/2007-2013)/ERC grant agreement No. 615187-COMANCHE, by the Marie Curie Initial Training Action (ITN) Q-NET 264034 and by the Marie Curie Individual Fellowship MSCA-IF-EF-ST No.~660532-SupeMag. The work of F.S.B was supported by Spanish Ministerio de Econom\' ia y Competitividad (MINECO) through the Project  FIS2014-55987-P and  the Basque Government under UPV/EHU Project No. IT-756-13.


\begin{thebibliography}{43}
\bibitem [{ber()}]{bergeret_footntefisnfis_2014}%
  \BibitemOpen
  \href@noop {} {}\bibinfo {note} {In principle at {$h_{ex}=\Delta_0/\sqrt{2}$}
  a first order transition from the superconducting to the normal state takes
  place. The solutions of the self-consistency equations for {$h_{ex}$} larger
  than this value exist as metastable states, and these are taken into account
  in our figures.}\BibitemShut {Stop}%
\end{thebibliography}
\end{document}